\documentclass[
reprint,
superscriptaddress,
showpacs,
 amsmath,amssymb,
 aps,
]{revtex4-1}

\usepackage{graphicx}% Include figure files
\usepackage{dcolumn}% Align table columns on decimal point
\usepackage{bm}% bold math
\usepackage{amsmath}
\usepackage{autobreak}
\usepackage{hyperref}% add hypertext capabilities
\usepackage[mathlines]{lineno}% Enable numbering of text and display math
\usepackage{natbib}
\usepackage{multirow}
\usepackage{mathtools}
\usepackage{color}
\usepackage{booktabs}
\usepackage{soul, color, xcolor}
\usepackage{amsthm,amssymb}
\usepackage{mathrsfs}
\usepackage{amssymb}
\usepackage{inconsolata}
\usepackage{url}
\usepackage{esvect}

\renewcommand{\vec}{\vv}

\makeatletter

\newcommand{\Rmnum}[1]{\expandafter\@slowromancap\romannumeral #1@}
\makeatother

\begin{document}

\preprint{APS/123-QED}

\title{Constraints on dark matter boosted by supernova shock within the effective field theory framework from the CDEX-10 experiment}

\author{J.~Z.~Wang}
\affiliation{Key Laboratory of Particle and Radiation Imaging (Ministry of Education) and Department of Engineering Physics, Tsinghua University, Beijing 100084}
\author{L.~T.~Yang}\altaffiliation [Corresponding author: ]{yanglt@mail.tsinghua.edu.cn}
\affiliation{Key Laboratory of Particle and Radiation Imaging (Ministry of Education) and Department of Engineering Physics, Tsinghua University, Beijing 100084}
\author{Q. Yue}\altaffiliation [Corresponding author: ]{yueq@mail.tsinghua.edu.cn}
\affiliation{Key Laboratory of Particle and Radiation Imaging (Ministry of Education) and Department of Engineering Physics, Tsinghua University, Beijing 100084}

\author{K.~J.~Kang}
\affiliation{Key Laboratory of Particle and Radiation Imaging (Ministry of Education) and Department of Engineering Physics, Tsinghua University, Beijing 100084}
\author{Y.~J.~Li}
\affiliation{Key Laboratory of Particle and Radiation Imaging (Ministry of Education) and Department of Engineering Physics, Tsinghua University, Beijing 100084}

\author{H.~P.~An}
\affiliation{Key Laboratory of Particle and Radiation Imaging (Ministry of Education) and Department of Engineering Physics, Tsinghua University, Beijing 100084}
\affiliation{Department of Physics, Tsinghua University, Beijing 100084}

\author{Greeshma~C.}
\altaffiliation{Participating as a member of TEXONO Collaboration}
\affiliation{Institute of Physics, Academia Sinica, Taipei 11529}

\author{J.~P.~Chang}
\affiliation{NUCTECH Company, Beijing 100084}
\author{H.~Chen}
\affiliation{Key Laboratory of Particle and Radiation Imaging (Ministry of Education) and Department of Engineering Physics, Tsinghua University, Beijing 100084}

\author{Y.~H.~Chen}
\affiliation{YaLong River Hydropower Development Company, Chengdu 610051}
\author{J.~P.~Cheng}
\affiliation{Key Laboratory of Particle and Radiation Imaging (Ministry of Education) and Department of Engineering Physics, Tsinghua University, Beijing 100084}
\affiliation{School of Physics and Astronomy, Beijing Normal University, Beijing 100875}
\author{W.~H.~Dai}
\affiliation{Key Laboratory of Particle and Radiation Imaging (Ministry of Education) and Department of Engineering Physics, Tsinghua University, Beijing 100084}
\author{Z.~Deng}
\affiliation{Key Laboratory of Particle and Radiation Imaging (Ministry of Education) and Department of Engineering Physics, Tsinghua University, Beijing 100084}
\author{C.~H.~Fang}
\affiliation{College of Physics, Sichuan University, Chengdu 610065}
\author{X.~P.~Geng}
\affiliation{Key Laboratory of Particle and Radiation Imaging (Ministry of Education) and Department of Engineering Physics, Tsinghua University, Beijing 100084}
\author{H.~Gong}
\affiliation{Key Laboratory of Particle and Radiation Imaging (Ministry of Education) and Department of Engineering Physics, Tsinghua University, Beijing 100084}
\author{Q.~J.~Guo}
\affiliation{School of Physics, Peking University, Beijing 100871}
\author{T.~Guo}
\affiliation{Key Laboratory of Particle and Radiation Imaging (Ministry of Education) and Department of Engineering Physics, Tsinghua University, Beijing 100084}
\author{X.~Y.~Guo}
\affiliation{YaLong River Hydropower Development Company, Chengdu 610051}
\author{L.~He}
\affiliation{NUCTECH Company, Beijing 100084}
\author{J.~R.~He}
\affiliation{YaLong River Hydropower Development Company, Chengdu 610051}

\author{H.~X.~Huang}
\affiliation{Department of Nuclear Physics, China Institute of Atomic Energy, Beijing 102413}
\author{T.~C.~Huang}
\affiliation{Sino-French Institute of Nuclear and Technology, Sun Yat-sen University, Zhuhai 519082}

\author{S.~Karmakar}
\altaffiliation{Participating as a member of TEXONO Collaboration}
\affiliation{Institute of Physics, Academia Sinica, Taipei 11529}

\author{H.~B.~Li}
\altaffiliation{Participating as a member of TEXONO Collaboration}
\affiliation{Institute of Physics, Academia Sinica, Taipei 11529}
\author{H.~Y.~Li}
\affiliation{College of Physics, Sichuan University, Chengdu 610065}
\author{J.~M.~Li}
\affiliation{Key Laboratory of Particle and Radiation Imaging (Ministry of Education) and Department of Engineering Physics, Tsinghua University, Beijing 100084}
\author{J.~Li}
\affiliation{Key Laboratory of Particle and Radiation Imaging (Ministry of Education) and Department of Engineering Physics, Tsinghua University, Beijing 100084}
\author{M.~C.~Li}
\affiliation{YaLong River Hydropower Development Company, Chengdu 610051}
\author{Q.~Y.~Li}
\affiliation{College of Physics, Sichuan University, Chengdu 610065}
\author{R.~M.~J.~Li}
\affiliation{College of Physics, Sichuan University, Chengdu 610065}
\author{X.~Q.~Li}
\affiliation{School of Physics, Nankai University, Tianjin 300071}
\author{Y.~L.~Li}
\affiliation{Key Laboratory of Particle and Radiation Imaging (Ministry of Education) and Department of Engineering Physics, Tsinghua University, Beijing 100084}
\author{Y.~F.~Liang}
\affiliation{Key Laboratory of Particle and Radiation Imaging (Ministry of Education) and Department of Engineering Physics, Tsinghua University, Beijing 100084}
\author{B.~Liao}
\affiliation{School of Physics and Astronomy, Beijing Normal University, Beijing 100875}
\author{F.~K.~Lin}
\altaffiliation{Participating as a member of TEXONO Collaboration}
\affiliation{Institute of Physics, Academia Sinica, Taipei 11529}
\author{S.~T.~Lin}
\affiliation{College of Physics, Sichuan University, Chengdu 610065}
\author{J.~X.~Liu}
\affiliation{Key Laboratory of Particle and Radiation Imaging (Ministry of Education) and Department of Engineering Physics, Tsinghua University, Beijing 100084}
\author{R.~Z.~Liu}
\affiliation{Key Laboratory of Particle and Radiation Imaging (Ministry of Education) and Department of Engineering Physics, Tsinghua University, Beijing 100084}
\author{S.~K.~Liu}
\affiliation{College of Physics, Sichuan University, Chengdu 610065}
\author{Y.~D.~Liu}
\affiliation{School of Physics and Astronomy, Beijing Normal University, Beijing 100875}
\author{Y.~Liu}
\affiliation{College of Physics, Sichuan University, Chengdu 610065}
\author{Y.~Y.~Liu}
\affiliation{School of Physics and Astronomy, Beijing Normal University, Beijing 100875}
\author{H.~Ma}
\affiliation{Key Laboratory of Particle and Radiation Imaging (Ministry of Education) and Department of Engineering Physics, Tsinghua University, Beijing 100084}
\author{Y.~C.~Mao}
\affiliation{School of Physics, Peking University, Beijing 100871}
\author{A.~Mureed}
\affiliation{College of Physics, Sichuan University, Chengdu 610065}
\author{Q.~Y.~Nie}
\affiliation{Key Laboratory of Particle and Radiation Imaging (Ministry of Education) and Department of Engineering Physics, Tsinghua University, Beijing 100084}
\author{H.~Pan}
\affiliation{NUCTECH Company, Beijing 100084}
\author{N.~C.~Qi}
\affiliation{YaLong River Hydropower Development Company, Chengdu 610051}
\author{J.~Ren}
\affiliation{Department of Nuclear Physics, China Institute of Atomic Energy, Beijing 102413}
\author{X.~C.~Ruan}
\affiliation{Department of Nuclear Physics, China Institute of Atomic Energy, Beijing 102413}
\author{M.~B.~Shen}
\affiliation{YaLong River Hydropower Development Company, Chengdu 610051}
\author{H.~Y.~Shi}
\affiliation{College of Physics, Sichuan University, Chengdu 610065}
\author{M.~K.~Singh}
\altaffiliation{Participating as a member of TEXONO Collaboration}
\affiliation{Institute of Physics, Academia Sinica, Taipei 11529}
\affiliation{Department of Physics, Banaras Hindu University, Varanasi 221005}
\author{T.~X.~Sun}
\affiliation{School of Physics and Astronomy, Beijing Normal University, Beijing 100875}
\author{W.~L.~Sun}
\affiliation{YaLong River Hydropower Development Company, Chengdu 610051}
\author{C.~J.~Tang}
\affiliation{College of Physics, Sichuan University, Chengdu 610065}
\author{Y.~Tian}
\affiliation{Key Laboratory of Particle and Radiation Imaging (Ministry of Education) and Department of Engineering Physics, Tsinghua University, Beijing 100084}
\author{H.~F.~Wan}
\affiliation{Key Laboratory of Particle and Radiation Imaging (Ministry of Education) and Department of Engineering Physics, Tsinghua University, Beijing 100084}
\author{G.~F.~Wang}
\affiliation{School of Physics and Astronomy, Beijing Normal University, Beijing 100875}

\author{L.~Wang}
\affiliation{School of Physics and Astronomy, Beijing Normal University, Beijing 100875}
\author{Q.~Wang}
\affiliation{College of Physics, Sichuan University, Chengdu 610065}
\author{Q.~Wang}
\affiliation{Key Laboratory of Particle and Radiation Imaging (Ministry of Education) and Department of Engineering Physics, Tsinghua University, Beijing 100084}
\affiliation{Department of Physics, Tsinghua University, Beijing 100084}
\author{Y.~F.~Wang}
\affiliation{Key Laboratory of Particle and Radiation Imaging (Ministry of Education) and Department of Engineering Physics, Tsinghua University, Beijing 100084}
\author{Y.~X.~Wang}
\affiliation{School of Physics, Peking University, Beijing 100871}
\author{H.~T.~Wong}
\altaffiliation{Participating as a member of TEXONO Collaboration}
\affiliation{Institute of Physics, Academia Sinica, Taipei 11529}

\author{Y.~C.~Wu}
\affiliation{Key Laboratory of Particle and Radiation Imaging (Ministry of Education) and Department of Engineering Physics, Tsinghua University, Beijing 100084}
\author{H.~Y.~Xing}
\affiliation{College of Physics, Sichuan University, Chengdu 610065}
\author{K.~Z.~Xiong}
\affiliation{YaLong River Hydropower Development Company, Chengdu 610051}
\author{R. Xu}
\affiliation{Key Laboratory of Particle and Radiation Imaging (Ministry of Education) and Department of Engineering Physics, Tsinghua University, Beijing 100084}
\author{Y.~Xu}
\affiliation{School of Physics, Nankai University, Tianjin 300071}
\author{T.~Xue}
\affiliation{Key Laboratory of Particle and Radiation Imaging (Ministry of Education) and Department of Engineering Physics, Tsinghua University, Beijing 100084}
\author{Y.~L.~Yan}
\affiliation{College of Physics, Sichuan University, Chengdu 610065}
\author{N.~Yi}
\affiliation{Key Laboratory of Particle and Radiation Imaging (Ministry of Education) and Department of Engineering Physics, Tsinghua University, Beijing 100084}
\author{C.~X.~Yu}
\affiliation{School of Physics, Nankai University, Tianjin 300071}
\author{H.~J.~Yu}
\affiliation{NUCTECH Company, Beijing 100084}
\author{X.~Yu}
\affiliation{Key Laboratory of Particle and Radiation Imaging (Ministry of Education) and Department of Engineering Physics, Tsinghua University, Beijing 100084}
\author{M.~Zeng}
\affiliation{Key Laboratory of Particle and Radiation Imaging (Ministry of Education) and Department of Engineering Physics, Tsinghua University, Beijing 100084}
\author{Z.~Zeng}
\affiliation{Key Laboratory of Particle and Radiation Imaging (Ministry of Education) and Department of Engineering Physics, Tsinghua University, Beijing 100084}

\author{F.~S.~Zhang}
\affiliation{School of Physics and Astronomy, Beijing Normal University, Beijing 100875}

\author{P.~Zhang}
\affiliation{YaLong River Hydropower Development Company, Chengdu 610051}
\author{Z.~H.~Zhang}
\affiliation{Key Laboratory of Particle and Radiation Imaging (Ministry of Education) and Department of Engineering Physics, Tsinghua University, Beijing 100084}
\author{Z.~Y.~Zhang}
\affiliation{Key Laboratory of Particle and Radiation Imaging (Ministry of Education) and Department of Engineering Physics, Tsinghua University, Beijing 100084}

\author{M.~G.~Zhao}
\affiliation{School of Physics, Nankai University, Tianjin 300071}

\author{J.~F.~Zhou}
\affiliation{YaLong River Hydropower Development Company, Chengdu 610051}
\author{Z.~Y.~Zhou}
\affiliation{Department of Nuclear Physics, China Institute of Atomic Energy, Beijing 102413}
\author{J.~J.~Zhu}
\affiliation{College of Physics, Sichuan University, Chengdu 610065}

\collaboration{CDEX Collaboration}
\noaffiliation

\date{\today}

\begin{abstract} 
Supernova shocks can boost dark matter (DM) particles to high, yet nonrelativistic, velocities, providing a suitable mechanism for analysis within the framework of the nonrelativistic effective field theory (NREFT). These accelerated DM sources extend the experimental ability to scan the parameter space of light DM into the sub-GeV region. In this study, we specifically analyze DM accelerated by the Monogem Ring supernova remnant, whose age ($\sim 68000$ yr) and distance to Earth ($\sim 300$ parsec) are strategically matched to enable detection with current terrestrial detectors. Utilizing the 205.4 kg$\cdot$day data obtained from the CDEX-10 experiment at the China Jinping Underground Laboratory, we derive new constraints on boosted DM within the NREFT framework. The NREFT coupling constant exclusion regions now penetrate the sub-GeV mass range, with optimal sensitivity achieved for operators $\mathcal{O}_{3}$, $\mathcal{O}_{6}$, $\mathcal{O}_{15}$ in the 0.4--0.6 GeV mass range.
\end{abstract}

\maketitle

\section{Introduction} 
Convincing evidence from both astrophysical observations and cosmological studies supports the existence of dark matter (DM, $\chi$)~\cite{PDG2023}, which accounts for approximately 26.8\% of the universe's energy budget~\cite{DarkMatterRatio}. Among the various DM candidates, weakly interacting massive particles (WIMPs) remain one of the most compelling. Extensive experimental efforts have been devoted to the direct detection (DD) of WIMPs through nuclear recoil signals, including XENON~\cite{XENON}, LUX~\cite{LUX}, PandaX~\cite{PandaX}, DarkSide~\cite{DarkSide}, CRESST~\cite{CRESST}, SuperCDMS~\cite{SuperCDMS}, CoGeNT~\cite{CoGeNT}, and CDEX~\cite{CDEX1,CDEX2,CDEX3,CDEX4,CDEX5,CDEX6,CDEX7,CDEX8,CDEXMigdal,CDEXEFT,CDEXCosmicRay}. However, to date, no experiment has observed a conclusive DM signal. This persistent null result continues to make dark matter one of the most profound mysteries in modern physics.

Traditional DD experiments conduct searches for DM through spin-independent (SI) and spin-dependent (SD) elastic scattering with ordinary nucleons ($\chi$-N). These experiments often rely on the standard halo model (SHM), which assumes that DM velocities follow a Maxwell-Boltzmann distribution with a local standard of rest velocity of 238 km/s and an escape velocity cutoff of 544 km/s~\cite{SHM_new}. However, light DM particles in the sub-GeV mass range remain undetectable within this conventional framework due to insufficient momentum transfer to overcome detector energy thresholds of current technologies. To address this limitation, various novel methodologies have been emerged to enhance DD sensitivity to lower DM mass regimes. For instance, inelastic scattering mechanisms, such as the Migdal effect, can extend the parameter space into the $m_\chi \sim$ $\mathcal{O}$(100 MeV) region ~\cite{CDEXMigdal,Migdal1}. Another promising strategy involves investigating boosted DM with higher momentum. In this context, potential sources of acceleration include high-energy cosmic rays ~\cite{CDEXCosmicRay,CosmicRay1,CosmicRay2,CosmicRay3,CosmicRay4,CosmicRay5,CosmicRay6}, blazars~\cite{Blazars1, Blazars2}, neutrinos~\cite{neutrino1, neutrino2, neutrino3, neutrino4}, the Sun~\cite{Solar1, Solar2, Solar3, Solar4, SolarCDEX}, and black halos~\cite{BlackHole1, BlackHoleCDEX}, etc. These (semi)relativistic DM particles enable current DD experiments to explore parameter space as low as $m_\chi \sim$ $\mathcal{O}$(10 keV), remarkably extending the discovery potential beyond traditional approaches.

Recently, supernova shocks have been proposed as a novel source of boosted DM. In this scenario, DM particles are accelerated through collisions with high-velocity nuclei within supernova remnants~\cite{SupernovaDM}, achieving speeds exceeding $0.01c$, an order of magnitude greater than typical DM velocities predicted by the SHM. In the case of ultralight DM particles, they can attain maximum velocities up to double the supernova shock speed through elastic scattering. This enhanced, yet nonrelativistic, velocity regime ($v\lesssim0.1c$) establishes supernova shock acceleration as a particularly well-suited mechanism for probing dark matter-nucleon interactions within the nonrelativistic effective field theory (NREFT) framework~\cite{EFT2013, EFT2014}, which systematically parametrizes $\chi$-N interactions through fourteen distinct operators. Besides conventional velocity-independent SI and SD scattering models, the NREFT architecture also encompasses numerous velocity-dependent operators. Under the supernova shock boost mechanism, the sensitivity of velocity-dependent interactions may increase by several magnitudes over that under SHM predictions, thereby substantially expanding the investigable parameter space for the corresponding operators.

The detectability of supernova shock accelerated DM critically depends on the progenitor supernova remnant's spatiotemporal characteristics. Only supernova remnant with appropriate age matching its distance to Earth could be an ideal candidate, providing currently observable DM fluxes. In this work, the Monogem Ring remnant~\cite{MonogemRing} emerges as an optimal candidate, fulfilling this temporal-spatial coincidence criterion. Based on the 205.4 kg$\cdot$day exposure data from the CDEX-10 experiment~\cite{C10sheze}, which employs $p$-type point contact high-purity germanium (PPCGe) detectors at the China Jinping Underground Laboratory (CJPL)~\cite{CJPL1, CJPL2}, we derive a set of constraints on NREFT operators. Our analysis incorporates simulation of Earth shielding effects~\cite{ESE1, ESE2, ESE3, ESE4, ESE5} from CJPL's  2400 m rock overburden through a modified version of the \texttt{CJPL\_ESS} simulation package~\cite{CDEXESE} developed by CDEX Collaboration.

\section{Effective field theory}

\begin{table}
\centering
\caption{Complete set of NREFT operators governing DM-nucleus interactions, with their corresponding cross section velocity scaling in the form of $\sigma_{N} \propto v^{2\alpha}$. Here $\alpha$ denotes the total power of momentum transfer $q$ and relative velocity $v$ in each operator's structure. Notably, $\mathcal{O}_{2}$ is typically excluded, as it cannot be derived from the leading-order nonrelativistic reduction of relativistic operators in effective field theory frameworks~\cite{EFT2014}.}
\begin{tabular}{p{1.5cm}<{\centering}p{4.5cm}<{\centering}p{2cm}<{\centering}}
\hline
\hline Operator & Formula & $v$-scale of $\sigma_{N}$ \\
\hline 
$\mathcal{O}_{1}$ & $1_{\chi}1_{N}$ & 0 \\
$\mathcal{O}_{2}$ & $(\vec{v}^{\perp})^2$ & ... \\
$\mathcal{O}_{3}$ & $i\vec{S}_{N}\cdot (\frac{\vec{q}}{m_{N}}\times \vec{v}^{\perp}_{N})$ & 4 \\
$\mathcal{O}_{4}$ & $\vec{S}_{\chi} \cdot \vec{S}_{N}$ & 0 \\
$\mathcal{O}_{5}$ & $i\vec{S}_{\chi}\cdot (\frac{\vec{q}}{m_{N}}\times \vec{v}^{\perp}_{N})$ & 4 \\
$\mathcal{O}_{6}$ & $(\vec{S}_{\chi}\cdot \frac{\vec{q}}{m_{N}})(\vec{S}_{N}\cdot \frac{\vec{q}}{m_{N}})$ & 4 \\
$\mathcal{O}_{7}$ & $\vec{S}_{N}\cdot \vec{v}^{\perp}_{N}$ & 2 \\
$\mathcal{O}_{8}$ & $\vec{S}_{\chi}\cdot \vec{v}^{\perp}_{N}$ & 2 \\
$\mathcal{O}_{9}$ & $i\vec{S}_{N}\cdot (\vec{S}_{N} \times \frac{\vec{q}}{m_{N}})$ & 2 \\
$\mathcal{O}_{10}$ & $i\vec{S}_{N}\cdot \frac{\vec{q}}{m_{N}}$ & 2 \\
$\mathcal{O}_{11}$ & $i\vec{S}_{\chi}\cdot \frac{\vec{q}}{m_{N}}$ & 2 \\
$\mathcal{O}_{12}$ & $\vec{S}_{\chi}\cdot (\vec{S}_{N} \times \vec{v}^{\perp}_{N})$ & 2 \\
$\mathcal{O}_{13}$ & $i(\vec{S}_{\chi}\cdot \vec{v}^{\perp}_{N})(\vec{S}_{N}\cdot \frac{\vec{q}}{m_{N}})$ & 4 \\
$\mathcal{O}_{14}$ & $i(\vec{S}_{\chi}\cdot \frac{\vec{q}}{m_{N}})(\vec{S}_{N}\cdot \vec{v}^{\perp}_{N})$ & 4 \\
$\mathcal{O}_{15}$ & $-(\vec{S}_{\chi}\cdot \frac{\vec{q}}{m_{N}})((\vec{S}_{N}\times \vec{v}^{\perp}_{N})\cdot\frac{\vec{q}}{m_{N}})$ & 6 \\
\hline
\end{tabular}
\label{tab:operators}
\end{table}

The NREFT provides a model-independent framework for describing $\chi$-N scattering processes. This approach systematically expands the effective Lagrangian in powers of momentum transfer($q$), where operators are truncated at leading order and next-to-leading-order to ensure computational tractability~\cite{EFT2013}. Within this formalism, all possible $\chi$-N interactions can be parametrized through linear combinations of four fundamental Galilean-invariant quantities:
\begin{equation}
i\vv{q}, \quad \vv{v}^{\perp}, \quad \vv{S}_{\chi}, \quad \vv{S}_{N}
\end{equation}
Here, $\vec{S}_{\chi}$ and $\vec{S}_{N}$ denote the spins operators of the DM particle and nucleon, respectively; $\vec{q}$ represents the momentum transfer vector during scattering; and $\vec{v}^{\perp}$ represents the transverse relative velocity between DM and nucleon. The NREFT framework initially derives 11 fundamental operators from these four Galilean invariants~\cite{EFT2013}, with Ref.~\cite{EFT2014} extending this to 15 operators by incorporating interactions mediated by higher-spin (beyond spin-0 or spin-1) fields. Each operator $\mathcal{O}_{i}$ is weighted by the coupling constants $c_{i}^{0}$ (isoscalar) and $c_{i}^{1}$ (isovector), reflecting the distinct nuclear response to DM interactions. One can calculate the differential cross section of $\chi$-N scattering with formula Eq.~\ref{eq:differentialCrossSection}:
\begin{align}
	&\frac{\text{d}\sigma_{N} (v,E_R)}{\text{d}E_R}=\frac{2m_T}{4\pi v^2}\ \left (\frac{1}{2j_{\chi}+1}\frac{1}{2j_{N}+1} \sum |\mathcal{M}|^2\right )\label{eq:differentialCrossSection}\\
	\notag
	&=\sum_{k} \sum_{\tau =0,1} \sum_{\tau'=0,1}R_k \left(\vec{v}^{\perp 2},\frac{\vv{q}^{2}}{m_N^2},\{ c_i^{\tau}c_j^{\tau'} \} \right)W_k^{\tau \tau'}(\vv{q}^{2} b^2).
\end{align}
The differential cross section depends on the relative velocity $v$ between DM and target nucleon, as well as the energy transfer $E_R$ during the collision. Here, $m_T$ denotes the mass of the target nucleus, while $j_{\chi}$ and $j_N$ represent the spins of the DM particle and nucleus, respectively. The EFT Galilean-invariant amplitude $|\mathcal{M}|$ is the product of WIMP and nuclear matrix elements. It is averaged over initial-state WIMP as well as nuclear magnetic quantum numbers ($M_{\chi}$, $M_N$) and summed over final magnetic quantum numbers. The index $k$ designates eight distinct WIMP-nucleon interaction types: $M$, $\Delta$, $\Sigma'$, $\Sigma''$, $\Phi'$ and $\Phi''$, which transform as vector charge, vector transverse magnetic, axial transverse electric, axial longitudinal, vector transverse electric, and vector longitudinal operators respectively, appending two interference terms $\Phi''M$ and $\Delta\Sigma'$. The WIMP response functions $R_k$ and nuclear response functions $W_k$ encapsulate the dynamics of these interactions. The coupling constants $c_i^{\tau}$ are embedded in $R_k$, directly modulating the differential cross section. Conversely, $W_k$ characterizes nuclear-specific properties (where $b$ indicates the nuclear size~\cite{EFT2013}) and varies across target nuclei. This work will only focus on isoscalar interactions and the superscript $^{0}$ of coupling constant $c_{i}^{0}$ will be omitted in later paragraphs, ignoring the isovector  $c_{i}^{1}$.

Crucially, NREFT operators exhibit a pronounced momentum dependence $(\sim q^{n})$, contrasting with conventional SI ($\mathcal{O}_{1}$) and SD ($\mathcal{O}_{4}$) cross sections that lack such scaling. The velocity scaling of the cross section $\sigma_{N}$ follows $\sigma_{N} \propto v^{2\alpha}$, where $\alpha$ corresponds to the combined power of momentum $q$ and velocity $v$ in the operator's analytic form (Table~\ref{tab:operators}). 
Under the SHM with typical DM velocities ($\mathcal{O}$(100 km/s)), SI/SD interactions dominate due to their velocity-independent nature. However, in scenarios with boosted DM velocities (e.g. $\mathcal{O}$($10^3$--$10^4$ km/s)), which remaining nonrelativistic, higher-order velocity-dependent operators may experience cross section enhancements of orders of magnitudes, potentially surpassing SI/SD contributions.

\section{DM Boosted By Supernova Ejecta}
During the expansion of a supernova shock wave, the initial stellar ejecta propagates outward and sweeps up the surrounding interstellar medium (ISM). The early evolution (first 100--200 yr) constitutes the free expansion phase (or ejecta-dominated phase), characterized by the swept-up ISM mass being negligible compared to the stellar ejecta mass. The transition to the Sedov-Taylor phase~\cite{SNR} occurs once the mass of the ambient matter swept up by the remnant exceeds the mass of the stellar ejecta. During this transition the shock begins to decelerate significantly. This phase is governed by the Sedov-Taylor solution, which expresses the shock expansion radius and velocity as functions of time since explosion, on the basis of the explosion energy $E_{\rm SN}$, ejecta mass $M_{\rm ej}$, and the ambient ISM density $n_{0}$. According to Refs.~\cite{S-Tfunction1, S-Tfunction2}, the shock radius is obtained as
\begin{equation}
R_{s}(t)=R_{0}\left(\left(\frac{t}{t_{0}}\right)^{-5\lambda_{\rm FE}}+\left(\frac{t}{t_0}\right)^{-5\lambda_{\rm ST}}\right)^{-1/5} \label{eq:Rs}
\end{equation}
Meanwhile, the shock velocity is derived as
\begin{align}
V_{s}(t)=&\frac{R_{0}}{t_{0}}\left(\frac{R_{s}(t)}{R_{0}}\right)^{6} \label{eq:Vs}\\
\notag
&\times\left(\lambda_{\rm FE}\left(\frac{t}{t_{0}}\right)^{-5\lambda_{\rm FE}-1}+\lambda_{\rm ST}\left(\frac{t}{t_{0}}\right)^{-5\lambda_{\rm ST}-1}\right),
\end{align}
where the scaling parameters $\lambda_{\rm FE}$ (free expansion phase), $\lambda_{\rm ST}$ (Sedov-Taylor phase), characteristic radius $R_{0}$, and characteristic time $t_{0}$ exhibit distinct values depending on supernova types and circumstellar environments. These parameters vary depending on the type of supernova. For a type $ \rm\Rmnum{1}$a supernova expanding across a uniform ISM, $\lambda_{\rm ST}=2/5$, $\lambda_{\rm FE}=4/7$, $R_{0}=\left(\frac{3M_{\rm ej}}{4\pi mn_{0}}\right)^{1/3}$, and $t_{0}=\left(R_{0}\left(\frac{M_{\rm ej}mn_{0}}{0.38E_{\rm SN}^{2}}\right)^{1/7}\right)^{7/4}$, where $m$ denotes the mean mass of the ISM. Meanwhile, for a type $\rm\Rmnum{2}$ supernova, the shock expands through a dense wind structured by its progenitor star before reaching the ISM. In this case, $\lambda_{\rm ST}=2/3$, $\lambda_{\rm FE}=6/7$, $R_{0}=\frac{M_{\rm ej}V_{w}}{\dot{M}}$, and $t_{0}=\left(R_{0}\left(\frac{\dot{M}}{36\pi}\frac{(18M_{\rm ej})^{-5/2}}{(40E_{\rm SN})^{-3/2}}\left(\frac{40E_{\rm SN}}{18M_{\rm ej}}\right)^{-9/2}\right)^{1/7}\right)^{7/3}$, where $V_{w}$ represents the presupernova wind velocity and $\dot{M}$ denotes the mass loss rate. The density of the presupernova wind is expressed as~\cite{S-Tfunction2}
\begin{equation}
\rho(r)=\frac{\dot{M}}{4\pi V_{w}r^{2}} \label{eq:rho},
\end{equation}
with $V_{w}=10$ km/s, and $\dot{M}=10^{-5}M_{\odot}/$yr following Ref.~\cite{S-Tfunction1}.

The Monogem Ring, the investigated target in this work, exhibits an angular diameter of $25^{\circ}$ on the celestial sphere as one of the closest known supernova remnants to Earth. Comprehensive analyses combining Sedov-Taylor hydrodynamical modeling, x-ray observations, and Galactic cosmic-ray propagation simulations~\cite{MonogemRing1,MonogemRing2, MonogemRing3} constrain its key parameters: distance to Earth $D=300$ parsec, age ${\rm Age}=68000$ yr, explosion energy $E_{\rm SN}=8.38\times10^{50}$ erg, and surrounding ISM density $n_{0}=3.73\times 10^{-3}$ $\rm cm^{-3}$~\cite{MonogemRing}. In the analysis below, the Monogen Ring is treated as type $ \rm\Rmnum{2}$ supernova, with the mean mass of ISM $m=1.27m_p$ (mass of proton), and the mass of supernova eject $M_{\rm ej}=5M_{\odot}$ (mass of the Sun) according to the typical values in Ref.~\cite{S-Tfunction2}. All parameters used for Monogem Ring modeling are summarized in Table~\ref{tab:SupernovaParameters}.

\begin{table}
	\centering
	\caption{Parameters used for Monogem Ring modeling.}
\begin{tabular}{p{5cm}<{\centering}p{3cm}<{\centering}}
\hline
\hline Parameter & Value \\
\hline 
Supernova type & II\\
$\lambda_{\rm ST}$ & 2/3\\
$\lambda_{\rm FE}$ & 6/7\\
$m$ (Mean mass of ISM) & 1.27$m_p$\\
$D$ (Distance to Earth) & 300 parsec\\
${\rm Age}$ & 68000 yr\\
$E_{\rm SN}$ (Explosion energy) & $8.38\times 10^{50}$ erg\\
$n_0$ (Surrounding ISM density) & $3.73\times 10^{-3}~\mathrm{cm}^{-3}$\\
$M_{\rm ej}$ (Mass of supernova ejecta) & $5M_{\odot}$ \\
$V_{w}$ (Presupernova wind velocity) & 10 km/s \\
$\dot{M}$ (Mass loss rate) & $10^{-5}M_{\odot}/$yr \\
\hline
\end{tabular}
\label{tab:SupernovaParameters}
\end{table}

\begin{table}
\centering
\caption{Mass fractions of the most abundant nuclei in supernova ejecta, derived from the averaged results in Ref.~\cite{ejecta}.}
\begin{tabular}{p{3cm}<{\centering}p{3cm}<{\centering}}
\hline
\hline Nucleus & $f_{i}$\\
\hline 
$\rm ^{1}H$ & 0.493\\
$\rm ^{4}He$ & 0.35\\
$\rm ^{16}O$ & 0.1\\
$\rm ^{28}Si$ & 0.02\\
$\rm ^{12}C$ & 0.015\\
$\rm ^{56}Fe$ & 0.007\\
$\rm ^{20}Ne$ & 0.005\\
$\rm ^{24}Mg$ & 0.005\\
$\rm ^{32}S$ & 0.005\\
$\rm ^{14}N$ & 0.004\\
$\rm ^{23}Na$ & 0.0004\\
\hline
\end{tabular}
\label{tab:ejecta}
\end{table}

Accurate modeling of boosted DM flux distributions requires precise characterization of the nuclear composition of supernova ejecta, as the $\chi$-N scattering cross sections exhibits strong dependence on the species of nuclei. Table~\ref{tab:ejecta} summarizes the mass fractions of dominant nuclei in supernova ejecta, derived from the ensemble-averaged results of five simulations in Ref.~\cite{ejecta}. The dominant nuclei are hydrogen and helium, occupying approximately 90\% of the total ejecta.

In this analysis, the supernova ejecta is modeled as a thin spherical shell with time-dependent radius $R_s(t)$ and expansion velocity $V_s(t)$, governed by the dynamical equations Eqs.~\ref {eq:Rs} and~\ref {eq:Vs}, respectively. This approximation is supported by both the Sedov-Taylor solution~\cite{SedovModel}, wherein the ejecta mass becomes concentrated near the shock front, and the more recent Chevalier model~\cite{ChevaierModel} demonstrating that supernova remnants exhibit sharply defined density gradients at the ejecta-environment interface. Under this thin-shell approximation, the DM particles encounter rate with the ejecta shell is given by 
\begin{equation}
4 \pi R_{s}(t)^{2}\frac{\rho_{\chi}}{m_{\chi}}V_{s}(t),
\end{equation}
where $\rho_{\chi}=0.3$ $\rm GeV/cm^3$ represents the local DM density near the Monogem Ring. Considering the suppressed $\chi$-N scattering cross section, the probability of multiscattering can be neglected. Under this assumption together with thin-shell approximation, the probability of a DM particle being boosted by a high-velocity nucleus is formally expressed:
\begin{equation}
\sum_{i} (\frac{M_{\rm ej}f_{i}}{m_{i}}+4\pi \int_{0}^{R_s(t)} n(r)r^2 \text{d} r\delta_{i,1})\frac{1}{4 \pi R_s(t)^2}\sigma_{\chi i}.
\end{equation}
Here, $\frac{M_{\rm ej}f_{i}}{m_{i}}$ represents the number of nuclei of species $i$, and $n(r)$ denotes the density of the presupernova wind, derived from Eq.~\ref{eq:rho}. The integral quantifies nuclei swept up by the wind, and $\delta_{i,1}$ exclusively considers the hydrogen contribution. $\frac{1}{4 \pi R_s(t)^2}$ arises from flux dilution across the expanding shell surface. To derive the velocity-dependent DM velocity, generalize the cross section to its differential form $\frac{\text{d}\sigma_{\chi i}}{\text{d}v}$, related to energy differentials through:
\begin{equation}
\frac{\text{d}\sigma_{\chi i}}{\text{d}v}=m_{\chi}v\frac{\text{d}\sigma_{\chi i}}{\text{d}E}, \label{eq:differential}
\end{equation}
where $E=\frac{1}{2}m_{\chi}v^2$.The resultant flux of upscattered DM particles with velocity $v$ at scattering instant $t$ becomes:
\begin{align}
\notag
\Phi(v,t)=&\int \text{d}E\delta(E-\frac{1}{2}m_{\chi}v^2)\rho_{\chi}V_s(t) \\ 
&\times \sum_{i} (\frac{M_{\rm ej}f_{i}}{m_{i}}+4\pi \int_{0}^{R_s(t)} n(r)r^2\text{d} r\delta_{i,1})v\frac{\text{d}\sigma_{\chi i}}{\text{d}E}, \label{eq:flux}
\end{align}
where $V_s(t)$ represents the velocity of nuclei in the ejecta before the collision, and $v$ denotes the velocity of upscattered DM particles. The parameter $m_{\chi}$ in Eq.~\ref{eq:differential} is implicitly included in the formula for the differential cross section. 
For terrestrial detection, the DM velocity $v$ and upscattering time $t$ must satisfy $v=D/({\rm Age}-t)$. Applying temporal delta function constraints to Eq.~\ref{eq:flux}, we obtain Earth-arriving DM particles flux:
\begin{equation}
\Phi_{Earth}(v)=\frac{1}{4\pi D^2}\int \Phi(v,t)\delta(t-({\rm Age}-D/v))\text{d}t. \label{eq:phi_earth}
\end{equation}

\begin{figure}[!tbp]
	\centering
\includegraphics[width=\linewidth]{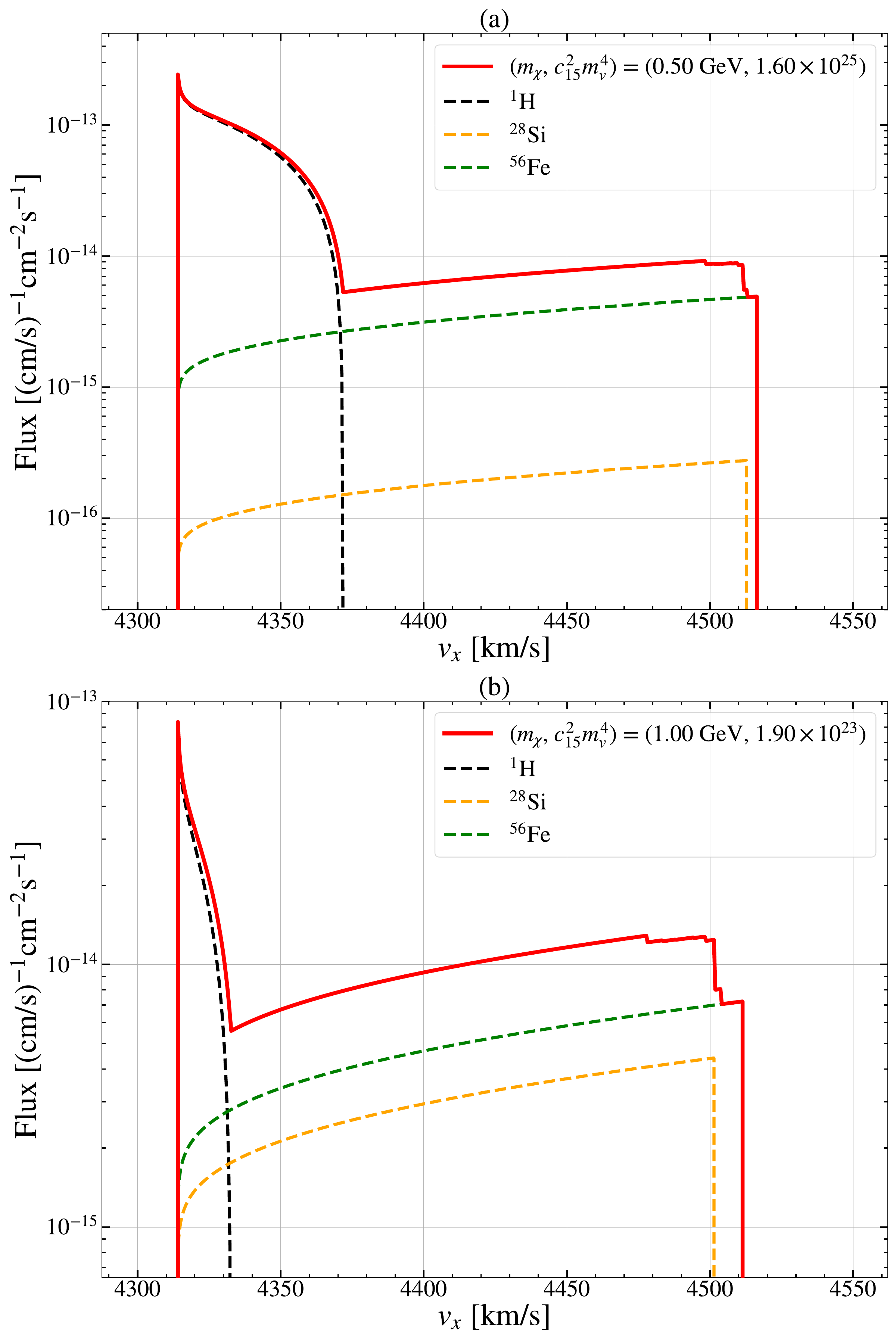}
\caption{Flux of boosted DM particles that can be detected on Earth, calculated using operator $\mathcal{O}_{15}$ with ($m_{\chi}, c_{15}^{2}m_{v}^{4}$) = (0.5 GeV, $1.6\times 10^{25}$) (a) as well as ($m_{\chi}, c_{15}^{2}m_{v}^{4}$) = (1.0 GeV, $1.9\times 10^{23}$) (b). Here, $m_v$ = 246 GeV denotes the weak mass scale.}
\label{fig:DM Flux Earth}
\end{figure}

The differential cross section formalism in the NREFT framework, as established in Ref.~\cite{EFT2013}, enables numerical computation of the boosted DM flux on Earth. To facilitate this analysis, we adapted the \texttt{Capt'n General}~\cite{Capt1, Capt2}, originally designed to analyze the solar DM capture within the NREFT framework, by refactoring its \texttt{Fortran} core into a \texttt{Python} implementation. Key parameters as well as equations governing the Monogem Ring’s shock dynamics were implemented.  The Earth-directed DM flux $\Phi_{\rm Earth}$ in Eq.~\ref{eq:phi_earth}, can be numerically evaluated through parametric inputs of DM mass $m_{\chi}$, selected operator $\mathcal{O}_{i}$ in Table~\ref{tab:operators}, and its corresponding coupling constant $c_i$. Figure~\ref{fig:DM Flux Earth} illustrates the computed terrestrial DM flux $\Phi_{\rm Earth}$ for ($m_{\chi}, c_{15}^{2}m_{v}^{4}$) = (0.5 GeV, $1.6\times 10^{25}$) as well as ($m_{\chi}, c_{15}^{2}m_{v}^{4}$) = (1.0 GeV, $1.9\times 10^{23}$), revealing velocity distribution features in 4300--4550 km/s range. The top three contributing nuclides, $\rm ^{1}H$, $\rm ^{28}Si$, and $\rm ^{56}Fe$ are displayed separately. The sharp peak in the low velocity region arises from hydrogen’s dominant abundance in the ejecta. Under the assumption of uniform ejecta expansion velocity, each nuclide generates a characteristic highest velocity edge, a consequence of elastic scattering kinematics where heavier nuclei impart greater momentum transfers to DM particles. The substantial contributions from $\rm ^{28}Si$ and $\rm ^{56}Fe$, despite their modest mass fractions, originate from the nuclide correlation in NREFT cross sections. Furthermore, the observed increase in flux with rising velocity further demonstrates the intrinsic correlation between interaction cross sections and momentum transfer dynamics within the NREFT framework.

\section{Data Analysis}
The recoil spectra for dark matter-nucleus elastic scattering in direct detection experiments are represented as
\begin{align}
\notag
\frac{\text{d}R}{\text{d}E_{R}}&=\frac{\rho_\chi}{m_{\chi}m_N}\int_{v_{min}(E_R)}^{\infty}vf(v)\frac{\text{d}\sigma_{\chi N}}{\text{d}E_R}d^3v \\
&=\frac{1}{m_N}\int_{v_{min}(E_R)}^{\infty}\Phi_{Earth}(v)\frac{\text{d}\sigma_{\chi N}}{\text{d}E_R}d^3v,
\end{align}
where $\Phi_{\rm Earth}$ (Eq.~\ref{eq:phi_earth}) corresponds to $\frac{\rho_{\chi}}{m_{\chi}}vf(v)$ in conventional analysis. In this work, the recoil spectra is computed utilizing the \texttt{WIMpy\_NREFT}~\cite{WIMpy} package, replacing its default Maxwell-Boltzmann velocity distribution $f(v)$ with our numerically derived $\Phi_{\rm Earth}(v)$ profile from Eq.~\ref{eq:phi_earth}.

In germanium semiconductor detectors, the detected energy, $E_{\rm det}$, relates to the actual nuclear recoil energy, $E_R$, owing to the quenching factor $Q_{nr}$, implying that $E_{det}=Q_{nr}E_R$~\cite{QF1,QF2,QF3}. In our analysis, the value of $Q_{nr}$ calculated using the \texttt{TRIM} package~\cite{TRIM} with a 10\% systematic error is utilized.

\begin{figure}[!tbp]
\centering
\includegraphics[width=\linewidth]{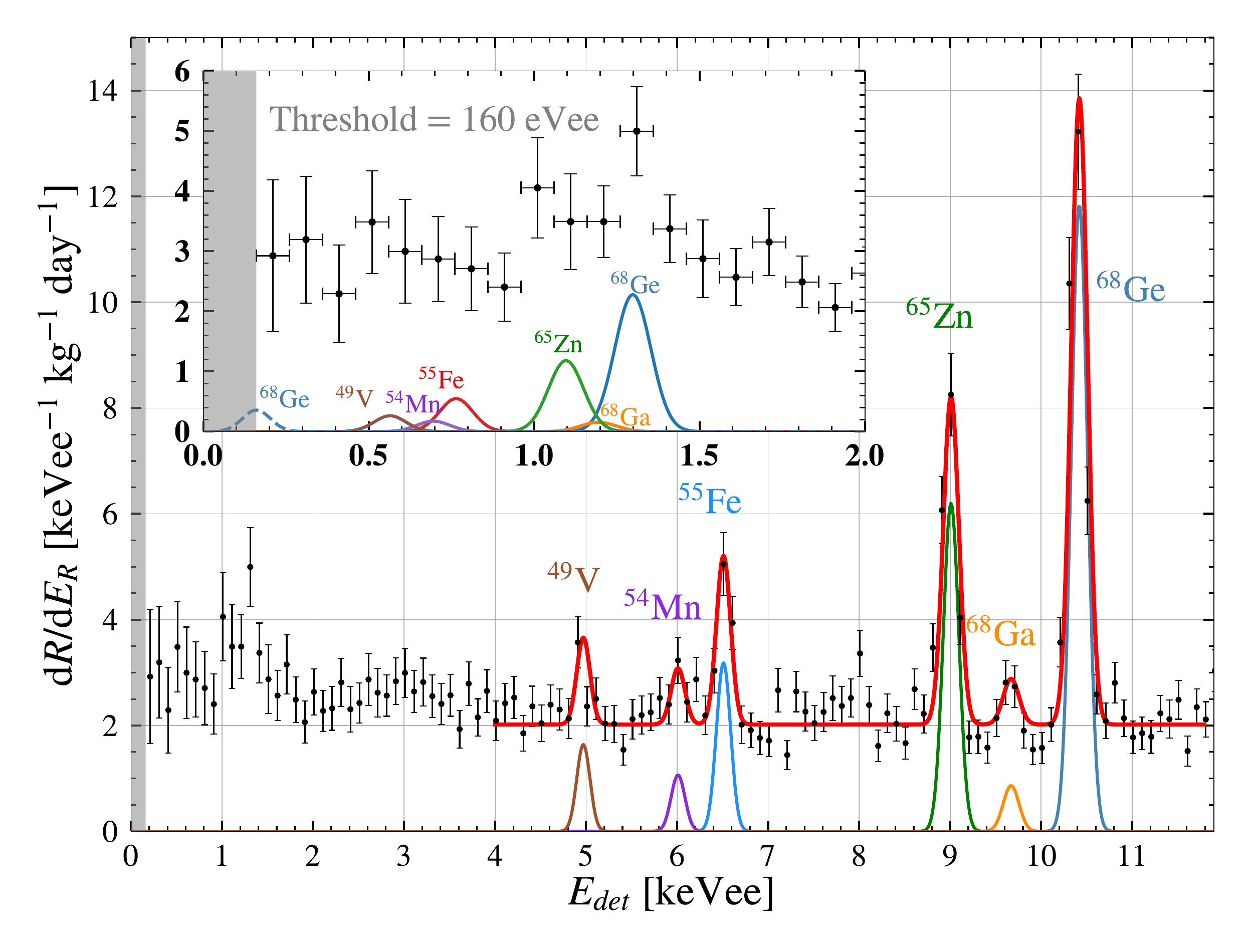}
\caption{Measured spectrum with error bars from the 205.4 kg$\cdot$day exposure data obtained from CDEX-10 in the energy range of 0.16--12 keVee. The red line represents the background model fit via $\chi^2$ minimization in the range of 4--11.8 keVee, including identified $K$-shell x-rays of cosmogenic radionuclides, displayed separately with other colors. The inset displays the contributions of $L$- and $M$-shell x-ray peaks, whose intensities are derived from corresponding $K$-shell lines.}
\label{fig:2}
\end{figure}

This study uses the 205.4 kg$\cdot$day exposure data obtained from the CDEX-10 experiment~\cite{C10sheze}. Previous studies have detailed the corresponding data processing procedures, including energy calibration, physics event selection, bulk-surface event discrimination, and a series of efficiency corrections~\cite{CDEX5,CDEX6,CDEX7,CDEX8}. The analysis threshold of CDEX-10 is 160 eVee, with a combined efficiency of 4.5\%~\cite{CDEX7}. Figure~\ref{fig:2} illustrates the final spectrum in the energy range of 0.16--12 keVee, along with fits for characteristic $K$-shell x-ray peaks from internal cosmogenic radionuclides such as $\rm ^{49}V$, $\rm ^{54}Mn$, $\rm ^{55}Fe$, $\rm ^{65}Zn$, $\rm ^{68}Ga$, and $\rm ^{68}Ge$. The inset displays the $L$-shell and $M$-shell x-ray peaks of these radionuclides, with the corresponding intensities derived from the $K$-shell peaks via fluorescence ratios~\cite{L/K}. The energy resolution of CDEX-10 is described by $\sigma(E_{\rm det})=35.8+16.6\times \sqrt{E_{\rm det}}$ (eV), where $E_{\rm det}$ is expressed in keV.

\begin{figure}[!tbp]
\centering
\includegraphics[width=\linewidth]{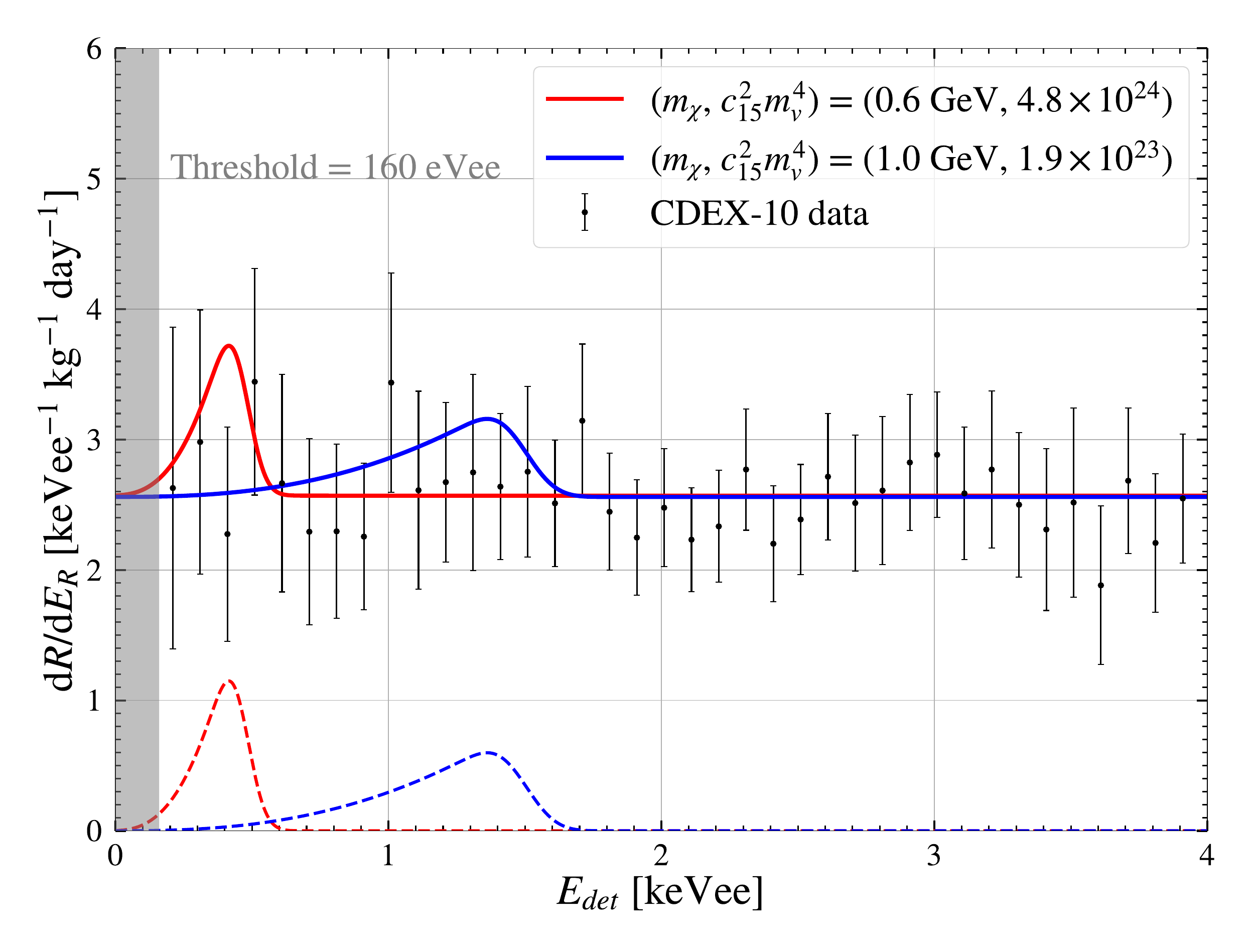}
\caption{Residual spectrum of CDEX-10 after subtracting the characteristic x-ray contributions in the energy region 0.16--4.00 keVee. The red and blue lines represent the predicted spectrum of supernova boosted DM for ($m_{\chi}, c_{15}^2m_{v}^{4}$) = (0.6 GeV, $4.80\times 10^{24}$) and ($m_{\chi}, c_{15}^2m_{v}^{4}$) = (1.0 GeV, $1.90\times 10^{23}$), where the coupling constants correspond to the upper limits at 90\% C.L. The dashed lines represent the expected signals deposited by boosted DM without background.}
\label{fig:3}
\end{figure}

After spectrum fitting and subtracting the contributions of characteristic x-rays, we employ the residual spectrum to determine the constraints on coupling constants via $\chi^2$ minimization~\cite{CDEX3}. The $\chi^2$ function is defined as
\begin{equation}
\chi^2(m_{\chi}, c_{i}^2m_{v}^{4})=\sum_{j=1}^{N}\frac{[n_j-B_j-S_j(m_{\chi}, c_i^2m_{v}^{4})]^2}{\sigma_{j}^2} \label{eq:chi2},
\end{equation}
where $n_j$ represents the measured event rate in the $j^{th}$ energy bin, $\sigma_j$ the total uncertainty incorporates both statistical and systematic uncertainties. The term $S_j(m_{\chi}, c_i^2m_{v}^{4})$ represents the predicted event rate for operator $\mathcal{O}_j$. The background component $B_{j}$ denotes the assumed background originates from the Compton scattering of high-energy $\gamma$ rays, modeled as a linear continuum $a\cdot E+b$. For each operator $\mathcal{O}_{i}$ and given $m_{\chi}$, the optimal $c_{i}^2m_{v}^{4}$ values are determined via $\chi^2$ minimization in the energy range of 0.16--12.00 keVee. Given that no significant DM signals are observed, the results are presented as upper limits on the coupling constants at the 90\% confidence level (C.L.), derived using the Feldman--Cousins method~\cite{FCmethod}. Figure~\ref{fig:3} illustrates the boosted DM spectrum corresponding to the upper limit at the 90\% C.L. for ($m_{\chi}, c_{15}^2m_{v}^{4}$) = (0.6 GeV, $4.80\times 10^{24}$) and ($m_{\chi}, c_{15}^2m_{v}^{4}$) = (1.0 GeV, $1.90\times 10^{23}$).

\begin{figure}[!tbp]
\centering
\includegraphics[width=\linewidth]{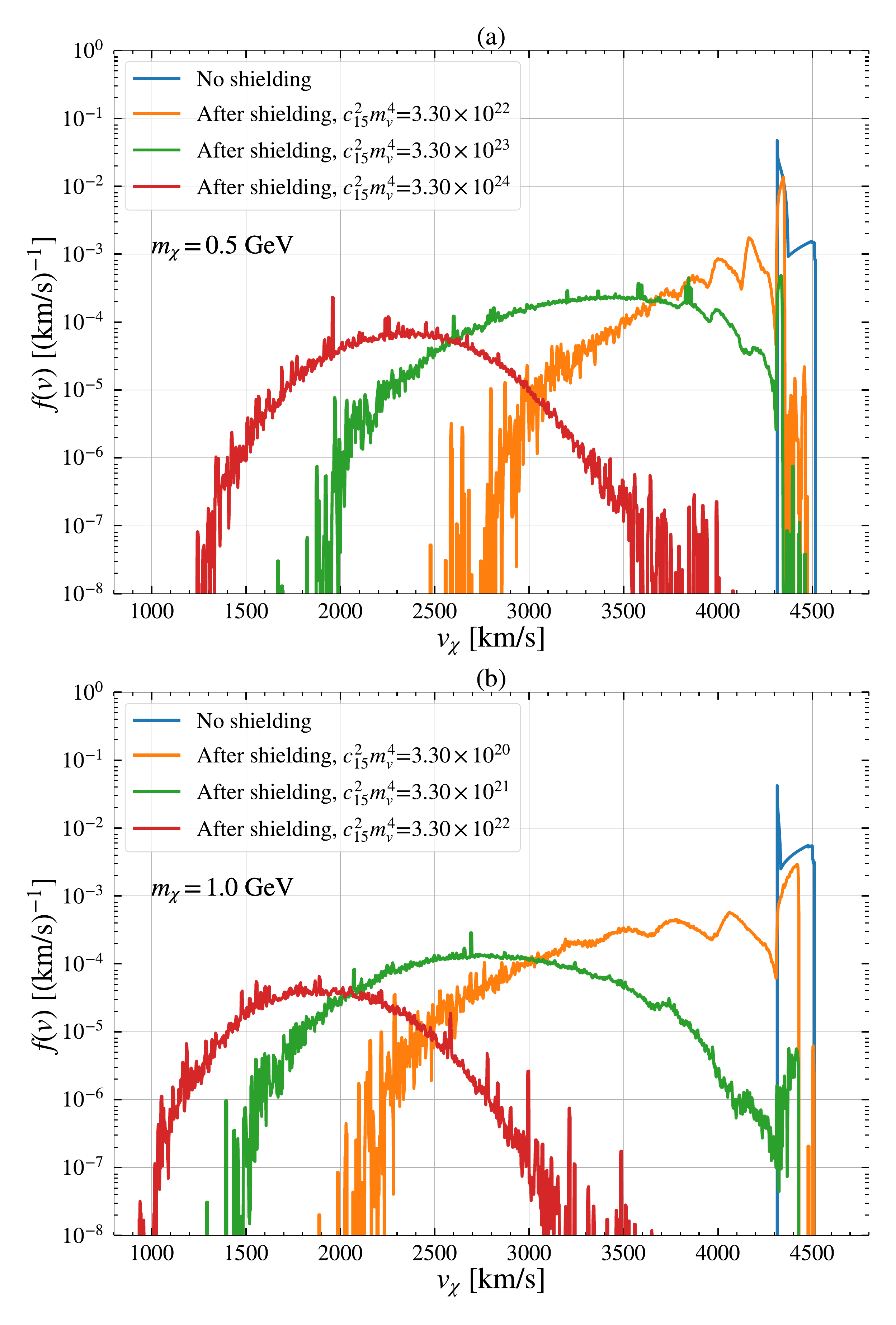}
\caption{Velocity distribution of 0.5 (a) and 1.0 GeV (b) DM under $\mathcal{O}_{15}$ interactions. The blue line represents the $f(v)$ of boosted DM reaching the Earth without shielding.}
\label{fig:4}
\end{figure}

The 2400-meter rock overburden at CJPL induces significant Earth attenuation of DM fluxes through $\chi$-N scattering, effectively decelerating particles, dispersing fluxes, and finally reducing detectable recoil energies. This is the so-called Earth shielding effect or Earth attenuation~\cite{ESE1, ESE2, ESE3, ESE4, ESE5}. To quantify this effect, a Monte Carlo simulation package \texttt{CJPL\_ESS}~\cite{CDEXESE} was developed by the CDEX Collaboration, in which a detailed geometric model and the rock compositions of Jinping Mountain are implemented. In this research, the package was upgraded by implanting the cross section formalism in the NREFT framework as well as incorporating the boosted DM source according to $\Phi_{\rm Earth}$ as defined in Eq.~\ref{eq:phi_earth}. Furthermore, to enhance simulation credibility, Earth's atmosphere is also incorporated into the \texttt{CJPL\_ESS}, whose effect can be significant within the exclusion region. The atmosphere is modeled with a thickness of 80 km, and the density profile across different altitudes complies with the U.S. Standard Atmosphere~\cite{USAtmosphere}. Figure~\ref{fig:4} displays simulation examples for the cases of $m_{\chi}$ = 0.5 and 1.0 GeV for operator $\mathcal{O}_{15}$. For larger coupling constants, the DM velocity distribution exhibits enhanced retardation after Earth shielding. 

\begin{figure*}[!htbp]
\includegraphics[width=\linewidth]{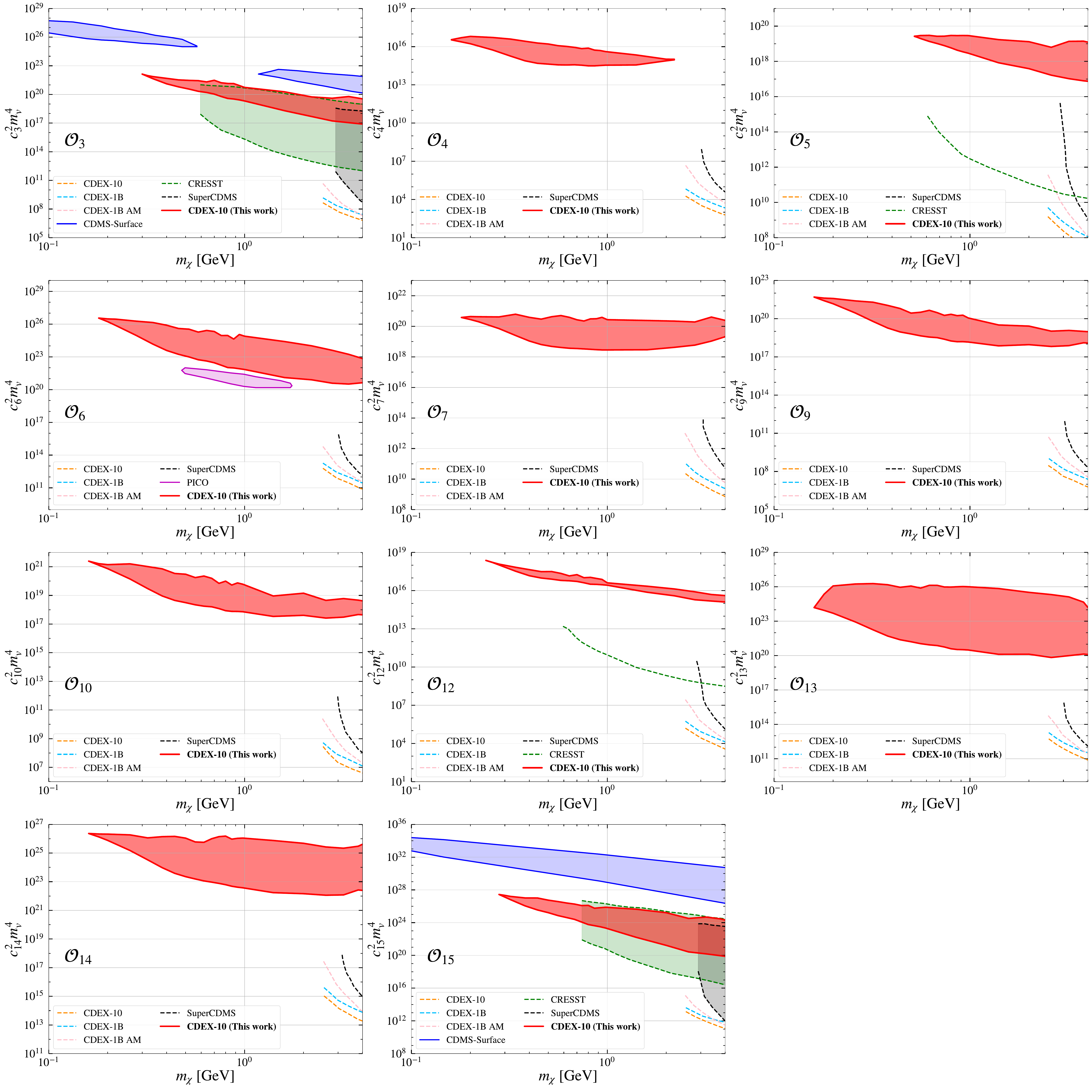}
\caption{Exclusion limits on NREFT coupling constants in the mass range of 0.1--4.0 GeV. Dashed lines correspond to constraints derived under the SHM scenario, including results from CRESST~\cite{CRESSTEFT}, SuperCDMS~\cite{SuperCDMSEFT}, CDEX-10, and CDEX-1B~\cite{CDEXEFT}, where ``CDEX-1B AM'' indicates the annual modulation (AM) analysis from the CDEX-1B experiment. Solid lines correspond to exclusion regions obtained from supernova shock boosted DM analyses. The constraints from CDMS-Surface~\cite{SuperCDMS-Surface} and PICO~\cite{PICO} experiments are detailed in Ref.~\cite{SupernovaDM}. The red lines indicate the 90\% confidence level exclusion limits for CDEX-10 derived in this work.}
\label{fig:5}
\end{figure*}

The exclusion regions at 90\% C.L. for supernova boosted DM are illustrated in Fig.~\ref{fig:5} as red lines. Here, the lower boundaries are derived using the minimal-$\chi^2$ method, while the upper boundaries are determined using the modified \texttt{CJPL\_ESS} simulations, incorporating Earth attenuations. Operators $\mathcal{O}_{1}$, $\mathcal{O}_{8}$, and $\mathcal{O}_{11}$ are excluded due to excessively large coupling constant values at lower boundaries, which preclude detectable energy deposition in \texttt{CJPL\_ESS} simulations and consequently prevent meaningful exclusion region determination. The dashed lines in Fig.~\ref{fig:5} represent published exclusion limits on NREFT coupling constants under the SHM scenario, as obtained from SuperCDMS~\cite{SuperCDMSEFT}, CRESST~\cite{CRESSTEFT}, CDEX-1B, and CDEX-10~\cite{CDEXEFT}. Other solid lines correspond to exclusion results for supernova boosted DM scenarios published by Ref.~\cite{SupernovaDM}, incorporating data from CDMS-Surface~\cite{SuperCDMS-Surface} and PICO~\cite{PICO} data for operators $\mathcal{O}_{3}$, $\mathcal{O}_{6}$, and $\mathcal{O}_{15}$. This investigation establishes the most stringent constraints to date for operators $\mathcal{O}_3$ and $\mathcal{O}_{15}$, in the mass range of 0.2--0.6 GeV. For other operators, the derived exclusion regions extend into previously unexplored sub-GeV parameter space, demonstrating novel coverage beyond existing experimental results.

\section{Discussion}
High-velocity nuclei in supernova shock fronts constitute a potent acceleration mechanism for DM particles. This investigation focuses on the ejecta of the Monogem Ring supernova remnant, characterized by substantial yet nonrelativistic expansion velocities ($v\lesssim0.1c$), rendering it particularly suitable for NREFT analysis. Since the exclusion results are highly depended on the parameters, we have also performed the analysis on type Ia supernova to examine its robustness. The discrepancy can be controlled under 4\%,  demonstrating the reliability of this approach.

The sensitivity of NREFT operators to specific detector materials imposes certain limitations on research. For instance, systems with spin-0 targets exhibit no sensitivity to purely spin-dependent operators such $\mathcal{O}_{6}$~\cite{SupernovaDM}. Furthermore, analyses become infeasible when the corresponding nuclear response functions remain unknown (e.g., tungsten~\cite{CRESSTEFT}). Detectors composed by nature germanium are not subject to these restrictions and maintain responsiveness across all operator scenarios, enabling comprehensive investigation of $\chi$--N scattering interactions and subsequent derivation of exclusion limits. Capitalizing on the pronounced cross-section dependence on momentum transfer inherent to the NREFT frameworks, our analysis achieves sub-GeV mass region with supernova boosted DM source. Operators exhibiting significant velocity scaling such as $\mathcal{O}_{3}$, $\mathcal{O}_{6}$, and $\mathcal{O}_{15}$ demonstrate particular sensitivity improvements. At higher mass ranges ($>$1 GeV), boosted DM exerts reduced efficacy compared to that under SHM in exclusion. This weakness originates from spherical diffusion effects during DM propagation from the supernova core to Earth, resulting in $\rho_{\chi}$ depletion spanning multiple orders of magnitude. 

Operators $\mathcal{O}_{1}$, $\mathcal{O}_{8}$, and $\mathcal{O}_{11}$ are excluded from effective constraint determination because the Earth shielding retardation is too strong at lower exclusion boundaries to collect detectable energy deposition in \texttt{CJPL\_ESS} simulation. This limitation is mainly due to dual mechanisms. Primarily, these operators exhibit limited velocity scaling characteristics (for example, $\mathcal{O}_{1}\sim v^{0}$ while $\mathcal{O}_{15}\sim v^{6}$), negating the velocity enhancement benefits inherent in the boosted DM system. Concurrently, the nuclear recoil cross sections in the NREFT framework exhibit significant dependence on target nuclide properties. The more enhanced $\chi$-N coupling under these particular operators in comparison with others induces substantial signal attenuation through the Earth shielding effect. For instance, the spin-0 nuclides in the mountain rock have no contribution to spin-dependent operator $\mathcal{O}_{4}$, yet they all make contributions to spin-independent operator $\mathcal{O}_{1}$. This dual mechanism, when combined with density depletion from diffusion attenuation of $\rho_{\chi}$, prevents detectable energy deposition in \texttt{CJPL\_ESS} simulations at coupling constants approaching lower exclusion boundaries. Consequently, no statistically significant exclusion boundaries could be established for these operators.

\section*{Acknowledgements}
This work was supported by the National Key Research and Development Program of China (Grants No. 2023YFA1607110 and No. 2022YFA1605000) and the National Natural Science Foundation of China (Grants No. 12322511 and No. 12175112). We would like to thank CJPL and its staff for hosting and supporting the CDEX project. CJPL is jointly operated by Tsinghua University and Yalong River Hydropower Development Company.

\bibliography{NREFT.bib}

\end{document}